\documentclass[preprint,prfluids,eqsecnum]{revtex4-2} % reprint longbibliography
\usepackage{natbib}
\usepackage{bm}
\usepackage{graphicx}
\usepackage{subfigure}
\usepackage{amssymb}
\usepackage{amsfonts}
\usepackage{amsmath}
\usepackage{amsbsy}
\usepackage{mathrsfs} 
\usepackage{color}
\usepackage{hyperref} 
\usepackage{soul}%\st{...}
\usepackage{mymacros}
\DeclareMathOperator\erf{erf}
\begin{document} 

\begin{abstract}
\end{abstract}

\title{Phoretic swimming with bulk absorption}
\author{Rodolfo Brand\~ao}
\affiliation{Department of Mechanical and Aerospace Engineering, Princeton University, Princeton, New Jersey 08544, USA}
\author{David Saintillan}
\affiliation{Department of Mechanical and Aerospace Engineering, University of California San Diego, La Jolla, California 92093, USA}
\author{Ehud Yariv}
\affiliation{Department of Mathematics, Technion --- Israel Institute of Technology, Haifa 32000, Israel}

\begin{abstract}
We consider phoretic self-propulsion of a chemically active colloid where solute is consumed at both the colloid boundary and within the bulk solution. Assuming first-order  kinetics, the dimensionless transport problem is governed by the surface Damk\"ohler number ${\mathcal{S}}$ and the bulk Damk\"ohler number ${\mathcal B}$. The dimensionless colloid velocity $U$, normalized by a self-phoretic scale,  is a nonlinear function of these two parameters. We identify two scenarios where these numbers are linked. When the controlling physical parameter is colloid size, ${\mathcal{S}}$ is proportional to ${\mathcal B}^{1/2}$; when the controlling parameter is solute diffusivity, ${\mathcal{S}}$ is proportional to ${\mathcal B}$. 
In the limit of small Damk\"ohler numbers, $U$ adopts the same asymptotic limit in both scenarios, proportional to ${\mathcal{S}}$. In the limit of large Damk\"ohler numbers, the deviations of solute concentration  from the equilibrium value are restricted to a narrow layer about the active portion of the colloid boundary. The asymptotic predictions of the associated boundary-layer problem are corroborated by an eigenfunction solution of the exact problem. The boundary-layer structure breaks down near the transition between the active and inactive portions of the boundary. The transport problem in that local region partially resembles the classical Sommerfeld  problem of wave diffraction from an edge.
%Both compared to numerical of exact. 
\end{abstract}

\date{\today}

\maketitle
\newpage

\date{\today}

\section{Introduction}
The remarkable propulsion exhibited by chemically active particles in liquid solutions, known as self-diffusiophoresis, has garnered significant attention following experimental breakthroughs in catalytic swimmers \cite{Aubret:17}. The fundamental mechanism underpinning phoretic self-propulsion involves two key components: solute  production or consumption at the particle boundary, coupled with short-range interactions between the solute molecules and that boundary. Golestanian \textit{et al.} \cite{Golestanian:07} introduced the first macroscale model to describe self-diffusiophoresis under Stokes flow conditions, accounting for diffusive solute transport. In that mode, chemical reactions at the particle boundary are represented through a prescribed solute flux distribution, while mechanical interactions with solute molecules are captured through a diffusio-osmotic slip velocity --- proportional to the tangential solute gradient at the outer edge of the interaction layer \cite{Anderson:89}. In the absence of solute advection, the linearity of the governing equations and boundary conditions implies that an asymmetry in the particle shape or physicochemical properties is required for self-propulsion: in typical experiments involving spherical colloids, this asymmetry is achieved by coating half of a particle with a catalyst. % responsible for the solute flux. %At finite P\'eclet numbers, a small perturbation in the solute distribution due to advection by the liquid can also generate a gradient around an isotropic colloid and induce a self-sustained translational motion by a linear instability \cite{Michelin:14}.

A more sophisticated model of surface reactions, which better describes experimental systems, involve first-order chemical kinetics \cite{Ebbens:12}. The associated boundary condition imposes a linear relation between the solute flux and local concentration, whose characteristic ratio defines the surface Dahmk\"ohler number (hereafter denoted by $\mathcal{S}$). For slow reaction rates ($\mathcal{S}\to 0$), the imposed flux model is recovered. Accounting for finite Dahmk\"ohler number has proven to be essential for capturing the dependence of the propulsion speed on particle size, as observed in experiments %, where the speed typically scales inversely with particle size 
\cite{Ebbens:12}. 

%Most of the modeling work on self-diffusiophoresis has considered particles in 3D bulk fluids, yet there has also been an interest in developing simpler analytical models in two dimensions. Such models may also be useful for analyzing various experiments in quasi-2D systems such as Hele-Shaw geometries. However, a mathematical conundrum arises in 2D in the absence of convection \cite{Sondak:16}, as the solution to Laplace's equation for the solute concentration around a particle acting as a net source or sink does not admit a solution that decays at infinity. This ill-posedness can be remedied by various modifications of the underlying model. For instance, a Janus particle that releases solute on one half and absorbs it by an equal amount on the other half induces not net source \cite{Crowdy:13}; such a restrictive assumption, however, is incompatible with most experimental systems which typically involve a net production of solute. Accounting for unsteadiness and/or finite solute advection \cite{Sondak:16,Yariv:17:2Dself,Yariv:20:self} can also cure the divergence of the concentration field at infinity and provide convergent expressions for the particle speed.

Of interest to us in this work is the case where the excess solute gets consumed in the bulk liquid surrounding the particle, for instance as a result of chemical degradation or bulk reaction with another solute. In that case, the strength of consumption is characterized by a bulk Dahmk\"ohler number, hereafter denoted by $\mathcal{B}$, defined as the ratio of the reactive to diffusive consumption rates.  Solute bulk absorption has already been studied using both numerical simulations \cite{deBuyl:13} as well as weakly nonlinear analyses near the threshold for spontaneous motion \cite{Schnitzer:23}. In certain ill-posed (e.g. steady self-phoresis in two dimensions
 \citep{Sondak:16,Yariv:17:2Dself}) and singular (e.g. spontaneous particle motion in channels \citep{Picella:22}) problems, even a weak bulk reaction may have significant effect.

Here, we analyze the steady motion of a a spherical self-phoretic particle. The paper is organized as follows. We formulate the problem in Sec.~\ref{sec:formulation} and present the dimensionless governing equations in Sec.~\ref{sec:dimensionless}. An exact solution based upon an eigenfunction expansion is derived in Sec.~\ref{sec:exact}. The linkage between the two Dahmk\"ohler numbers is discussed in Sec.~\ref{sec:smallD}, where we also consider the case of small $\mathcal S$ and $\mathcal B$. The limit of large Dahmk\"ohler numbers is addressed in Sec.~\ref{sec:largeD}. The associated boundary-layer analysis breaks down near the junction between the active and inactive portions of the particle boundary. We analyze the structure of this transition region in Sec.~\ref{sec:transition}. Illustrative examples are presented in Sec.~\ref{sec:numerics}. We conclude in Sec.~\ref{sec:conclusion}.

\section{Problem formulation} \label{sec:formulation}
A chemically active spherical  particle (radius $a$) is freely suspended in an unbounded  solution (solute diffusivity $D$). The equilibrium solute concentration, at large distances from the particle, is denoted by $c_\infty$. %instant centre $O$ from wall $(1+\delta)a$. 
Solute transfer at the particle boundary is %The particle is chemically reactive. Using  the chemical absorption interface 
modeled using a first-order chemical reaction \citep{Ebbens:12},
\begin{equation}
\text{solute absorption (per unit area)} = k \times \text{local solute concentration}, \label{fixed rate}
\end{equation}
where the (positive) rate constant $k$ generally varies along  the boundary. In addition, we assume \citep{Yoshinaga:12,deBuyl:13} that solute is consumed in the bulk in proportion to the excess concentration --- the deviation of its concentration from the equilibrium value,
\begin{equation}
\text{solute consumption (per unit volume)} = k_b \times \left\{\text{local solute concentration}-c_\infty\right\}. \label{bulk}
\end{equation}
The (positive) bulk rate $k_b$ is assumed uniform.

Following the prevailing practice \citep{Ebbens:12,Michelin:14}, we restrict the subsequent analysis to situations where $k$ is  symmetric about an axis passing through the particle  center; self-propulsion accordingly takes place in the form of rigid translation in a direction parallel to that axis. Our goal is the associated speed. %acquired by the particle.

Defining $\bar k$ as a characteristic norm of $k$, relation \eqref{fixed rate} leads to the definition of the Damk\"ohler number,
\begin{equation}
{\mathcal{S}} = \frac{a\bar k}{D}, \label{Da}
\end{equation}
representing the ratio of reactive ($\bar k c_\infty$) to diffusive ($D c_\infty/a$) solute flux densities. %reaction to diffusion.
The problem is also affected by the bulk Damk\"ohler number,
\begin{equation}
{\mathcal B} = \frac{a^2k_b}{D}, \label{Da_b}
\end{equation}
representing the ratio of reactive ($k_b c_\infty$) to diffusive ($D c_\infty/a^2$) consumption rates.

We employ a macroscale description, where the short-range interaction between the solute molecules and the particle is manifested by
diffusio-osmotic slip \citep{Anderson:89},
\begin{equation}
\text{slip velocity} = b \times \text{surface gradient of solute concentration}. \label{diffusio-osmosis}
\end{equation}
We assume that $b$ is uniform. Note  that $b$ is a signed quantity, positive for repulsive  interactions and negative for attractive ones. %(e.g. steric) (e.g. van der Waals) onteractions. 
 The velocity scale associated with \eqref{diffusio-osmosis} is $\mathcal{U} = bc_\infty/a$. %{|b|\mathcal{J}}/{D}$. 
%This scale may be positive or negative, which is OK here.

\begin{figure}[hbtp]
\centering
\includegraphics[scale=0.4]{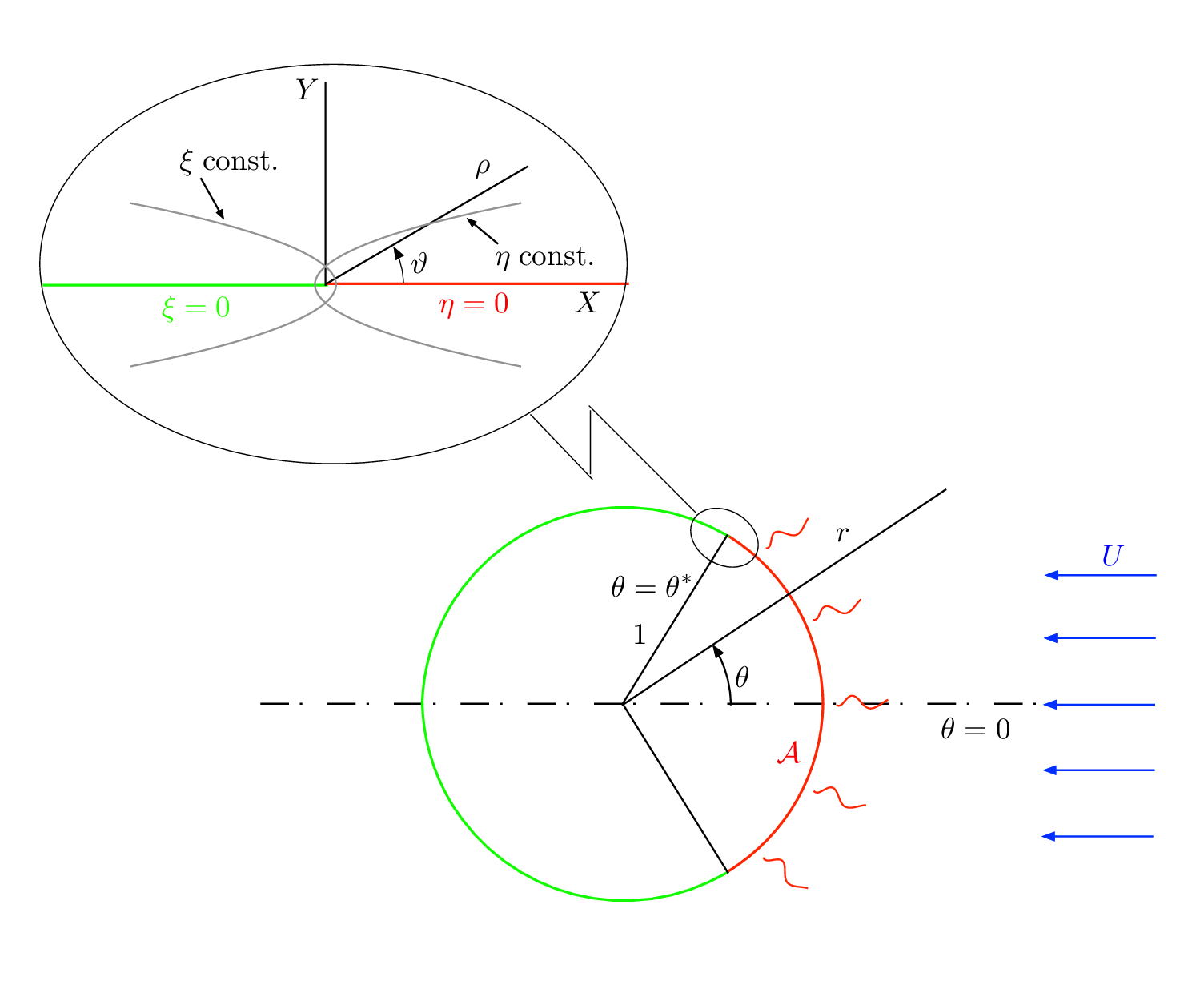} % .eps [scale=0.55]
\caption{Schematic showing the particle geometry and coordinates. The zoomed region (rotated) describes the transition-region coordinates.} 
\label{fig:schematic}
\end{figure}

We adopt a co-moving reference frame with the origin at the particle center. In that frame we utilize the spherical coordinates $(ar,\theta,\phi)$
defined such that the axis $\theta=0,\pi$ is aligned along the symmetry diameter of the particle and $r=1$ is the particle boundary, see Fig.~\ref{fig:schematic}. %The corresponding unit vectors are denoted $(\uniti,\unitj)$. 
The specified axially-symmetric activity is then represented by the  function $k(\theta)$.

Consistently with the macroscale description, the particle acquires the rectilinear velocity required to keep it force-free. Due to the axial symmetry,  the particle velocity relative to the otherwise quiescent liquid must be parallel to the symmetry axis, say $U^* \uniti$ ($\uniti$ being a unit vector in the direction $\theta=0$). 
Our goal is the calculation of $U^*$.
\section{Dimensionless description} \label{sec:dimensionless}
%We employ a dimensionless notation where all length variables are normalized by $a$, as well as a particle-fixed reference system with origin at the particle center. In that system we utilize the spherical coordinates $(r,\theta,\phi)$defined such that the axis $\theta=0,\pi$ is aligned along the symmetry diameter of the particle. %The corresponding unit vectors are denoted $(\uniti,\unitj)$. The axially-symmetric activity is represented by $k(\theta)$.

In what follows we consider the coupled transport--flow problem governing the excess solute concentration $c$, normalized by $c_\infty$, and fluid velocity $\bu$,
normalized by $\mathcal{U}$. Given the presumed axial symmetry, $c$ is a function of $r$ and $\theta$. In the particle-fixed reference frame, the velocity of the particle is manifested as the uniform streaming $-U \uniti$ at infinity, where $U=U^*/\mathcal U$.

The dimensionless %Following the macroscale description of \cite{Michelin:14}, 
solute transport  problem is governed by: (i) the diffusion--reaction equation,
\begin{equation}
\Laplacian c = {\mathcal B}c, \label{AD}
\end{equation}
wherein
\begin{equation}
\label{Laplacian}
\Laplacian = \pd{}{r^2} + \frac{2}{r}\pd{}{r}  + \frac{1}{r^2\sin^2\theta}\pd{^2}{\theta^2} 
%\bnabla = \be_r \pd{}{r} + \frac{\be_\theta}{r}\pd{}{\theta},
\end{equation}
is the pertinent dimensionless Laplacian; %gradient operator;
(ii) the kinetic condition at the particle boundary,
\begin{equation}
\pd{c}{r} =  {\mathcal{S}}  (1+c) f(\theta) \quad \text{at} \quad r=1, \label{imposed rate}
\end{equation}
where $f(\theta)=k(\theta)/\bar k$ is the dimensionless distribution of rate constant; and 
(iii) the approach to equilibrium at large distances,
\begin{equation}
c \to 0 \quad \text{as} \quad r\to\infty. \label{far c}
\end{equation}

Note that condition \eqref{imposed rate} is meaningful only with solute consumption, $f\geq 0$. The preceding problem provides $c$ as a function of the governing parameters ${\mathcal{S}}$ and ${\mathcal B}$, as well as $f(\theta)$. Once solved, we can consider the flow, governed by: (i) the continuity and Stokes equations [the former tacitly employed in \eqref{AD}]; (ii) the diffusio-osmotic slip [cf.~\eqref{diffusio-osmosis}]
\begin{equation}
\bu = \be_\theta \pd{c}{\theta} \quad \text{at} \quad r=1;
\label{slip}
\end{equation}
(iii) the far-field approach to a uniform stream (see Fig.~\ref{fig:schematic}),
\begin{equation}
\bu \to - U \uniti \quad \text{as} \quad r\to\infty; \label{far bu}
\end{equation}
and (iv) the requirement that the particle is force-free. 
In fact, the detailed calculation of the flow field is not required, as use of the reciprocal theorem \citep{Stone:96} provides $U$ as the quadrature
\begin{equation}
U = \frac{1}{2}\int_{0}^\pi \left.\pd{c}{\theta}\right|_{r=1} \sin^2\theta \, \dd \theta,
\label{Stone}
\end{equation}
or, following integration by parts, % provides $U$ as the quadrature
\begin{equation}
U = -\int_0^\pi \left.c\right|_{r=1} \sin\theta\cos\theta \, \dd \theta. \label{w as int of c}
\end{equation}

%%Of course, with $c$ affected by advection, \eqref{w as int of c} should be regarded as a useful relation rather than a formal solution of the problem. Thus, 
%We therefore see explicitly that $U$ is proportional to $M$.  %Since the present transport problem is ill-posed at $\Pen=0$, it is not \textit{a priori} evident whether the symmetry property \eqref{M symmetry} holds as $\Pen\to0$. 
In what follows, it may be convenient to employ $\mu=\cos\theta$ instead of $\theta$. 
Writing $f(\theta)=F(\mu)$, it is natural to represent $F$ as a series of surface harmonics,
\begin{equation}
\label{surface harmonics}
F(\mu) = \sum_{m=0}^\infty F_m P_m(\mu)
\end{equation}
wherein $P_m$ are the Legendre polynomials of degree $m$. Using the orthogonality
of these polynomials,
\begin{equation}
\int_{-1}^1 P_m(\mu)P_n(\mu) \, \dd\mu = \frac{2}{2m+1} \delta_{mn},
\label{orthogonality}
\end{equation}
 we obtain
\begin{equation}
F_m = \frac{2m+1}{2} \int_{-1}^1 F(\mu)  P_m(\mu)\, \dd\mu. \label{F_m}
\end{equation}
%\section{Activity profile}

When using $\mu$, \eqref{w as int of c} simplifies to %=\cos\theta$ instead of $\theta$. 
\begin{equation}
U = -\int_{-1}^1 \mu\left.c\right|_{r=1} \, \dd \mu .
 \label{w as int of c better}
\end{equation} 
\section{Exact solution} \label{sec:exact}
Using the eigenfunctions of the modified Helmholtz equation, we find that the most general axisymmetric solution of \eqref{AD}--\eqref{Laplacian} and \eqref{far c} is
\begin{equation}
c=\sum_{n=0}^\infty A_n r^{-1/2}K_{n+1/2}({\mathcal{B}}^{1/2}r)P_n(\cos\theta).
\label{eigenfunctions}
\end{equation}
Here $K_\nu$ are the modified Bessel functions of the second kind with degree $\nu$.
Substitution of \eqref{surface harmonics} and \eqref{eigenfunctions} into condition \eqref{imposed rate} yields
\begin{multline}
\sum_{n=0}^\infty A_n \left[ K_{n+1/2}'({\mathcal{B}}^{1/2}) - \frac{1}{2}
K_{n+1/2}({\mathcal{B}}^{1/2})\right]P_n(\mu) \\
= \mathcal{S} \left[1+ \sum_{n=0}^\infty A_n K_{n+1/2}({\mathcal{B}}^{1/2})P_n(\mu)\right]\sum_{m=0}^\infty F_m P_m(\mu), \label{BC in An}
\end{multline}
where the prime denote differentiation with respect to the argument.
Projection of \eqref{BC in An} upon $P_m(\mu)$ ($m=0,1,2,\ldots$) yields an infinite linear system governing the coefficients $\{A_n\}_{n=0}^\infty$. % are determined using condition\eqref{imposed rate}, whose projection upon $P_m(\mu)$ (where $m=0,1,2,\ldots$) yields an infinite linear system governing these coefficients. 
Using controlled truncation, this system may be solved in principle for any values of $\mathcal B$ and $\mathcal S$ and a given activity distribution $f(\theta)$. Once solved, substitution into \eqref{w as int of c better} yields, upon making use of the orthogonality relations \eqref{orthogonality},
\begin{equation}
U = -\frac{2}{3}A_1 K_{3/2}({\mathcal{B}}^{1/2}). \label{U exact}
\end{equation}

Prior to illustrating the exact solution for a specific activity distribution, it is desirable to 
supplement it by asymptotic approximations.
\section{Linked Damk\"ohler numbers} \label{sec:smallD}
Considering the manner by which the Damk\"ohler numbers \eqref{Da}--\eqref{Da_b}  depend upon the dimensional quantities in the problem, there are two natural scenarios where these two numbers are linked. The first scenario, 
\begin{equation}
{\mathcal{S}} \propto \sqrt{{\mathcal B}}, \label{prop1}
\end{equation}
corresponds to the situation where the particle size $a$ is allowed to vary, while 
all other dimensional quantities are fixed.   
The second scenario, 
\begin{equation}
{\mathcal{S}} \propto {\mathcal B}, \label{prop2}
\end{equation}
corresponds to the situation where it is the diffusivity $D$ that is allowed to vary. %all dimensional quantities, other than $D$, are fixed. 

These linkages suggest that in an asymptotic analysis, we should study the situation where  both numbers are either small or large. For small Damk\"ohler numbers, the leading-order calculation is actually independent of the linkage. 
Indeed, it is evident from \eqref{imposed rate} that $c$ is of order $\mathcal S$, while from 
\eqref{AD} we see that at leading-order $c$ is governed by Laplace's equation. Writing $c= \mathcal S \acute c + \cdots$, we find from \eqref{imposed rate} that $c'$ satisfies
\begin{equation}
\pd{\acute c}{r} =  f(\theta) \quad \text{at} \quad r=1. 
\label{imposed rate small B S}
\end{equation}

Writing the harmonic field  $\acute c$ as a sum of spherical harmonics, % solution of the form 
\begin{equation}
\acute c =  \sum_{m=0}^\infty a_m \frac{P_m(\mu)}{r^{m+1}} ,
\label{c1 expan}
\end{equation} 
we readily obtain using \eqref{surface harmonics},
\begin{equation}
a_m = -(m+1)^{-1}F_m \quad \text{for} \quad m\neq0. \label{a_n}
\end{equation} 
The particle velocity is then obtained from \eqref{orthogonality}, \eqref{w as int of c better} and \eqref{a_n}:
\begin{equation}
U = \frac{{\mathcal{S}}  F_1}{3}+ \cdots. \label{regular}
\end{equation}
This leading-order velocity is unaffected by bulk absorption.

\section{Large Damk\"ohler numbers} \label{sec:largeD}
In the limit of large Damk\"ohler numbers we find from \eqref{AD} and \eqref{far c} that
\begin{equation}
c\equiv0. \label{identically 0}
\end{equation}
Since this is an exact solution of both \eqref{AD} and \eqref{far c}, it is evident the asymptotic error is exponentially small.

The trivial solution \eqref{identically 0} is clearly incompatible with \eqref{imposed rate} at $\mathcal A$, the active portion of the boundary (see Fig.~\ref{fig:schematic}),  %is the active portion of the boundary, %[say in $(0,\pi)$]
\begin{equation}
\mathcal A  = \{\theta \in (0,\pi)| f(\theta)>0\}.
\end{equation} %$r=1$, where $f>0$. 
Seeking an additional distinguished limit at large Damk\"ohler numbers, we observe from \eqref{AD} a possible dominant balance with spatial variations across a narrow region of $\ord({\mathcal B}^{-1/2})$ width. We therefore postulate a boundary layer of that width about $\mathcal A$.
Defining the stretched coordinate
\begin{equation}
{Y} =  {\mathcal B}^{1/2} (r-1), \label{def rho}
\end{equation}
we write in the boundary layer
\begin{equation}
c(r,\theta;\mathcal B) = \tilde c({Y}, \theta;\mathcal B) . \label{layer expansion no Da} %\frac{}{\sqrt{{\mathcal B}}}
\end{equation}
Substitution of \eqref{def rho}--\eqref{layer expansion no Da} into the diffusion--reaction equation \eqref{AD} yields 
\begin{equation}
\pd{^2\tilde c}{{Y}^2} + {\mathcal B}^{-1/2}\pd{\tilde c}{{Y}} + \cdots = \tilde c
 \quad \text{for} \quad {Y}>0.
\label{AD layer}
\end{equation}
Condition \eqref{imposed rate} becomes,
\begin{equation}
{\mathcal B}^{1/2} \pd{\tilde c}{{Y}} =  {\mathcal{S}}  (1+\tilde c) f(\theta) \quad \text{at} \quad {Y}=0, \label{imposed rate layer}
\end{equation}
and the requirement of matching with the ``outer'' solution \eqref{identically 0}
implies the far-field decay %of $\tilde c$,
\begin{equation}
\lim_{{Y}\to\infty}\tilde c=0. \label{BL decay}
\end{equation}
Once the boundary-layer problem is solved, the particle speed is readily obtained from \eqref{w as int of c} as
\begin{equation}
U = -\int_{\mathcal A} \left.\tilde c\right|_{{Y}=0} \sin\theta\cos\theta \, \dd \theta.
\label{U large Da}
\end{equation} 

We only seek the leading-order solution. Thus, we have from \eqref{AD layer}
\begin{equation}
\pd{^2 \tilde c}{{Y}^2} = \tilde c\quad \text{for} \quad {Y}>0.
\label{Helmholtz 1D}
\end{equation}
The solution of \eqref{BL decay} and \eqref{Helmholtz 1D} is %that decays at large ${Y}$ is
\begin{equation}
\tilde c({Y},\theta) = -L(\theta)\ee^{-{Y}}. \label{BL exp}
\end{equation}
Substitution into \eqref{U large Da} gives
\begin{equation}
U = \int_{\mathcal A} L(\theta) \sin\theta\cos\theta \, \dd \theta.
\label{U large Da in A}
\end{equation} 

Up to this point, the analysis has been independent of the linkage between $\mathcal B$ and $\mathcal S$.
The distribution $L(\theta)$, however, is determined by condition \eqref{imposed rate layer}, whose leading-order form depends upon the specific linkage. The case \eqref{prop1} of linkage by size is
conveniently represented by the relation
\begin{equation}
{\mathcal{S}} = \alpha \sqrt{{\mathcal B}} , \label{linkage1}
\end{equation}
where $\alpha$ is fixed. Condition \eqref{imposed rate layer} then reads, at leading order,
\begin{equation}
\pd{\tilde c}{{Y}} =  \alpha (1+\tilde c) f(\theta) \quad \text{at} \quad {Y}=0. \label{imposed rate layer link1}
\end{equation}
Substitution of \eqref{BL exp} then gives
\begin{equation}
L(\theta) = \frac{\alpha f(\theta)}{1+ \alpha f(\theta)}. \label{A case I}
\end{equation}
We therefore find from \eqref{U large Da in A} that, at leading order,
\begin{equation}
U = \alpha\int_{\mathcal A}\frac{ f(\theta)}{1+ \alpha f(\theta)} \sin\theta\cos\theta \, \dd \theta.
\label{U large Da in f 1}
\end{equation} 

The case \eqref{prop2} of linkage by diffusivity is represented by the relation,
\begin{equation}
{\mathcal{S}} = \beta {{\mathcal B}} \label{linkage2}
\end{equation}
where $\beta$ is considered fixed.
Here, at leading order, condition \eqref{imposed rate layer}  gives
\begin{equation}
\tilde c=-1 \quad \text{at} \quad {Y}=0. \label{imposed rate layer link2}
\end{equation}
Substitution of \eqref{BL exp} then gives
\begin{equation}
L(\theta)\equiv1. \label{A is 1}
\end{equation}
We therefore find from \eqref{U large Da in A} that
\begin{equation}
U = \int_{\mathcal A}\sin\theta\cos\theta \, \dd \theta,
\label{U large Da in f 2}
\end{equation} 
at leading order.

Remarkably, the particle velocity depends only upon the active fraction of  boundary; it is independent of $\beta$ and $f$ and is accordingly insensitive to the details of the activity profile. In what follows, it is convenient to restrict the analysis to the case (see Fig.~\ref{fig:schematic}) where $\mathcal A=(0,\theta^*)$ with $0<\theta^*<\pi$ [cf.~\eqref{Janus} and \eqref{Janus gen}]. Under this modest restriction, \eqref{U large Da in f 2} gives
\begin{equation}
U = \frac{\sin^2\theta^*}{2}. \label{very general Janus}
\end{equation}

\section{Transition region} \label{sec:transition}
The solution in the limit of large Damk\"ohler numbers, with linkage by diffusivity, 
may appear to introduce a contradiction. Indeed, the nonzero velocity %\eqref{U large Da in f 2} 
\eqref{very general Janus}, which may be traced back to  \eqref{w as int of c}, is incompatible with the zero velocity predicted by a naive substitution of \eqref{BL exp} and \eqref{A is 1} into the original quadrature \eqref{Stone}.
The origin of this incompatibility  has to do with smoothness at the boundary $r=1$.  %The solution in limit of large Damk\"ohler numbers, with linkage by diffusivity, is not smooth at the surface $r=1$. 
Indeed, the excess concentration is discontinuous at the  transition $\theta=\theta^*$ between $\mathcal A$, about which \eqref{BL exp} and \eqref{A is 1} hold, and
its complement, about which \eqref{identically 0} holds. 
With a finite discontinuity, expression \eqref{Stone} cannot be applied in a piecewise manner.

The resolution of this apparent contradiction has to do with a breakdown of the boundary-layer structure. 
The boundary-layer solution, where variations with respect to $\theta$ are assumed ``small,'' is clearly incompatible with a finite discontinuity. A transition region is therefore formed about the edge ($r=1$ and $\theta =\theta^*$) of $\mathcal A$. In that region,  the excess concentration 
smoothly varies from the boundary-layer solution \eqref{BL exp} and \eqref{A is 1} at $\theta<\theta^*$ to the nil value \eqref{identically 0}
 at $\theta>\theta^*$. With the presence of such a region, the original quadrature \eqref{Stone} is dominated by a small neighborhood $\mathcal N$ of $\theta^*$, which is still asymptotically larger than the width of the transition region. Since $\theta$ is approximately constant in that neighborhood, we  obtain from \eqref{Stone}
\begin{equation}
U = \frac{\sin^2\theta^*}{2}\int_{\mathcal N} \left.\pd{c}{\theta}\right|_{r=1}  \, \dd \theta.
\label{Stone approx}
\end{equation}
Recalling the need to match the minus unity value for $\theta^*>\theta$ and the zero value for $\theta>\theta^*$, we retrieve \eqref{very general Janus}.

The boundary-layer scaling suggests that the lateral extent of the transition region is
$\mathcal B^{-1/2}$. Defining the local coordinate [cf.~\eqref{def rho}]
\begin{equation}
X = \mathcal B^{1/2}(\theta^*-\theta), 
\end{equation}
and considering the limit $\mathcal B\to\infty$ with $X,Y$ fixed we find that
the transition region coincides the upper half $XY$-plane (see Fig.~\ref{fig:schematic}). Defining $C(X,Y) = -c(r,\theta)$, $C$ is governed by the modified Helmholtz equation
\begin{equation}
\pd{^2C}{X^2}+\pd{^2C}{Y^2} = C \quad \text{for} \quad Y>0.\label{modified Helmholtz}
\end{equation}
At large $Y$ it must satisfy
\begin{equation}
\lim_{Y\to\infty}C=0, \label{transition decay}
\end{equation}
representing asymptotic matching with \eqref{identically 0}.

It remains to specify the mixed boundary conditions at $Y=0$, which follow from the exact condition \eqref{imposed rate layer} with linkage by diffusivity \eqref{linkage2}:
\begin{equation}
\pd{C}{{Y}} =  -\beta {\mathcal B}^{1/2} (1-C) f(\theta^* - {\mathcal B}^{-1/2}X) \quad \text{at} \quad {Y}=0. 
\label{active portion link1}
\end{equation}
In the inert portion of the boundary, where $f=0$, we find
\begin{equation}
\pd{C}{Y}=0 \quad \text{for} \quad X<0. \label{transition left BC}
\end{equation}
The condition on the active portion depends upon the asymptotic behavior of $f(\theta)$ as $\theta\nearrow\theta^*$. %, see \eqref{imposed rate layer}. 
In the case where $f(\theta)$ attains there a nonzero limit [cf.~\eqref{Janus}], 
applying the limit ${\mathcal B}\to\infty$ to \eqref{active portion link1} yields % we have
\begin{equation}
C=1 \quad \text{for} \quad X>0; \label{simple BC}
\end{equation}
in the case where $f(\theta) \sim K(\theta^*-\theta)$ as $\theta\nearrow\theta^*$ [cf.~\eqref{Janus gen}, where $K=1$], the appropriate condition in the limit ${\mathcal B}\to\infty$ is 
\begin{equation}
\pd{C}{Y} =  -\beta K (1-C) X \quad \text{for} \quad X>0. \label{difficult BC}
\end{equation}

For situations where \eqref{simple BC} holds, the problem is reminiscent of the diffraction of plane waves of sound by the edge of a semi-infinite screen --- a problem originally solved by Sommerfeld \cite{Sommerfeld:96}. 
Defining the local polar coordinates $(\rho,\vartheta)$ by
\begin{equation}
X = \rho \cos\vartheta, \quad Y =  \rho \sin\vartheta \label{polar}
\end{equation}
(see Fig.~\ref{fig:schematic}),
the solution of \eqref{modified Helmholtz}--\eqref{transition decay} and \eqref{transition left BC}--\eqref{simple BC}, derived in the Appendix, is
\begin{equation}
C = \frac{\ee^{-Y}}{2} \left\{ 1 + \erf \left[\rho^{1/2} \left( \cos\frac{\vartheta}{2}
-\sin\frac{\vartheta}{2}\right) \right]\right\} 
+ \frac{\ee^{Y}}{2} \left\{ 1 - \erf \left[\rho^{1/2} \left( \cos\frac{\vartheta}{2}
+\sin\frac{\vartheta}{2}\right) \right]\right\}. \label{transition final}
\end{equation}

%where the domain becomes $0<\vartheta<\pi$.
In terms of the polar coordinates \eqref{polar}, the limit $X\to\infty$ with $Y$ fixed corresponds to $\rho\to\infty$ with 
$\vartheta = O(1/\rho)$. We then readily obtain
\begin{equation}
\lim_{X\to\infty}C=\ee^{-Y},
\end{equation}
which trivially matches the boundary-layer solution  \eqref{BL exp} and \eqref{A is 1}.
The limit $X\to-\infty$ with $Y$ fixed corresponds to $\rho\to\infty$ with 
$\pi-\vartheta = O(1/\rho)$. Here, we obtain
\begin{equation}
\lim_{X\to-\infty}C=0,
\end{equation}
which trivially matches the nil concentration about the inert portion of the particle boundary.

\section{Illustrations} \label{sec:numerics}
\begin{figure}[hbtp]
\centering
\includegraphics[scale=0.4]{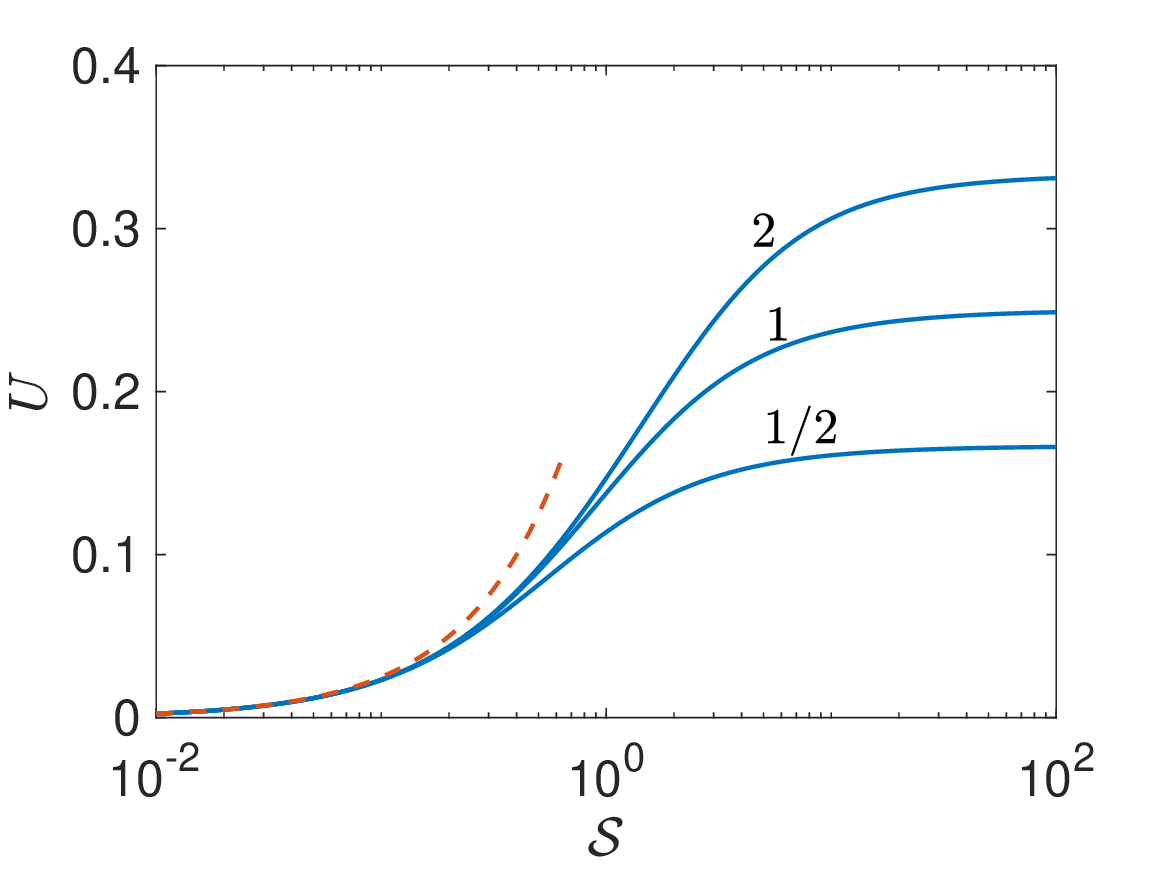} % .eps [scale=0.55]
\caption{$U$ versus ${\mathcal S}$ using linkage-by-size \eqref{linkage1} for the Janus profile \eqref{Janus}. Solid: exact result \eqref{U exact} for the indicated values of $\alpha$. Dashed: small-$\mathcal S$ approximation \eqref{regular}.} 
\label{fig:U}
\end{figure}

We continue by illustrating our results, considering first the case of linkage by size. 
With 
$\mathcal B$ locked to $\mathcal S$ via  \eqref{linkage1}, $U$ becomes a function of $\mathcal S$, $\alpha$ and the activity profile. We use a Janus configuration, namely
\begin{equation}
\label{Janus}
f(\theta) = \left\{ 
\begin{array}{ll}
  1, &   0 < \theta < {\pi}/{2}   ,    \\
  0, &   {\pi}/{2} < \theta < {\pi} ,
\end{array}
\right.
\end{equation}
for which \eqref{F_m} gives %$F_{2k} = \delta_{k,0}/2$ and $F_{2k+1} = (-)^k/(2k+1)\pi$.
\begin{equation}
F_{2k} = \frac{\delta_{k,0}}{2}, \quad F_{2k+1} = \frac{(-)^k(2k)! (4k+3)}{2^{2k+2}(k!)^2(k+1)}, \label{F Janus}
\end{equation}
and, in particular, $F_1=3/4$. The velocity calculated using \eqref{U exact}
is shown in Fig.~\ref{fig:U} for $\alpha=1/2$, $1$ and $2$. We also 
portray the $\alpha$-independent small Damk\"ohler-number approximation \eqref{regular}, which here gives $U = \mathcal{S}/{4}$ for $\mathcal{S}\ll1$. With \eqref{Janus}, the large Damk\"ohler-number approximation \eqref{U large Da in f 1}  gives 
 \begin{equation}
\lim_{\mathcal{S}\to\infty} U = \frac{\alpha}{2(1+\alpha)}. \label{large Da_b U linked Janus 1}
\end{equation}
For the aforementioned $\alpha$ values, it implies the respective limits $1/6$, $1/4$ and $1/3$.
The approach at large $\mathcal{S}$ to these limits is evident in the figure. 
%In Fig.~\ref{fig:U}. Note the approach at large $\mathcal S$ to the limit \eqref{large Da_b U linked Janus}, which for these $\alpha$ values respectively gives $1/6$, $1/4$ and $1/3$.

\begin{figure}[hbtp]
\centering
\includegraphics[scale=0.4]{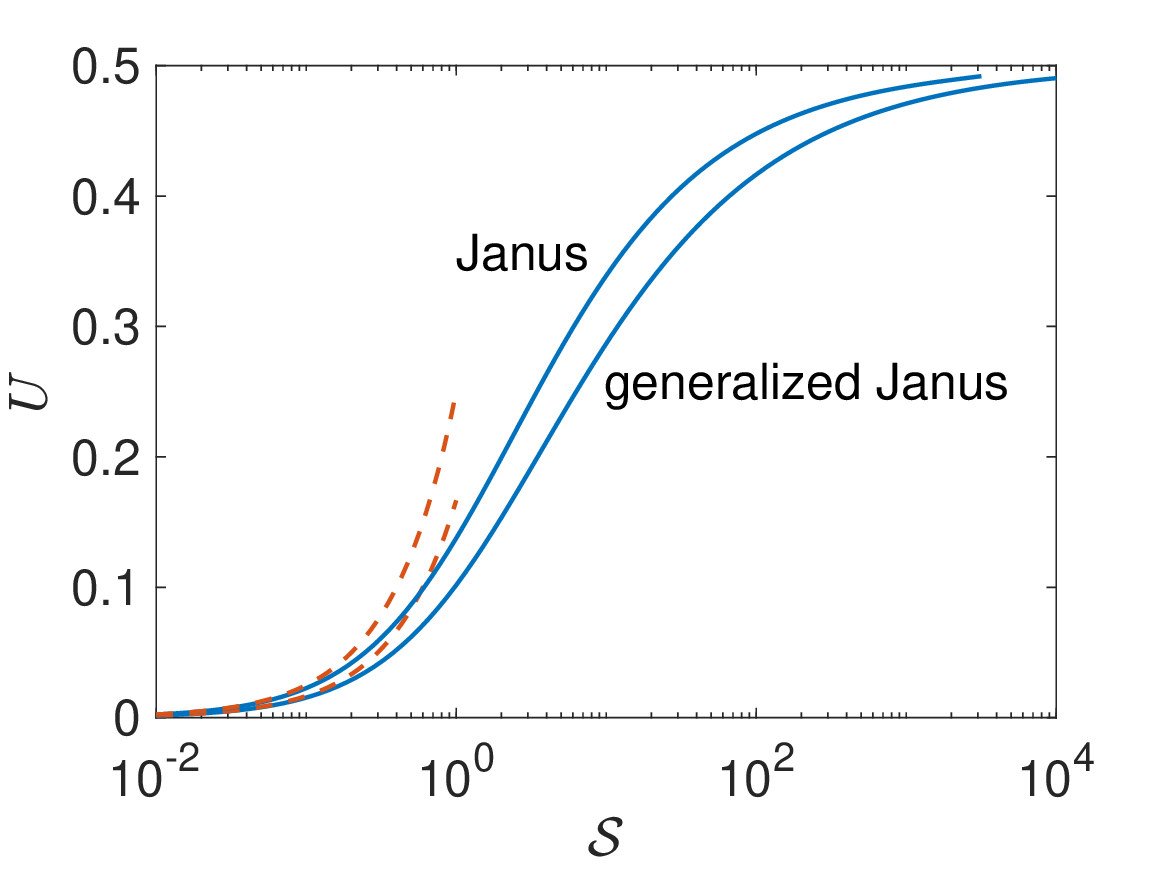} % .eps [scale=0.55]
\caption{$U$ versus ${\mathcal B}$ using linkage-by-diffusivity \eqref{linkage2} with $\beta=1$ for both the Janus \eqref{Janus} and the generalized Janus \eqref{Janus gen} activity profiles. Solid: exact result \eqref{U exact}. Dashed: small Damk\"ohler-number approximation \eqref{regular}.} 
\label{fig:U_linkage_by_D}
\end{figure}

Consider now the linkage by diffusivity. With $\mathcal B$ locked to $\mathcal S$ via   \eqref{linkage2}, $U$ becomes a function of $\mathcal S$, $\beta$ and the activity profile. We here use a single linkage value, $\beta=1$, but consider both the Janus  activity distribution \eqref{Janus} and the generalized Janus profile %[cf.~\eqref{Janus}]
\begin{equation}
\label{Janus gen}
f(\theta) = \left\{ 
\begin{array}{ll}
  \cos\theta, &   0 < \theta < {\pi}/{2}   ,    \\
  0, &   {\pi}/{2} < \theta < {\pi} ,
\end{array}
\right.
\end{equation}
for which %[cf.~\eqref{F Janus}]
\begin{equation}
F_{2k} = \frac{(-)^{k+1}(2k)!(4k+1)}{4^{k+1}(k!)^2(k+1)(2k-1)}, \quad F_{2k+1} = \frac{\delta_{k0}}{2},
\label{F Janus ge}
\end{equation}
and, in particular, $F_1=1/2$.
For that profile the small Damk\"ohler-number approximation  \eqref{regular} gives $U = \mathcal{S}/{6}$ for $\mathcal{S}\ll1$. Since ${\mathcal A}=(0,\pi/2)$ for both \eqref{Janus} and \eqref{Janus gen}, these distributions share the same large Damk\"ohler-number limit
\eqref{very general Janus}, namely 
 \begin{equation}
\lim_{\mathcal{S}\to\infty} U = \frac{1}{2}. \label{large Da_b U linked Janus 2}
\end{equation}
The results are illustrated in Fig.~\eqref{fig:U_linkage_by_D}. % we present $U$ versus ${\mathcal S}$ using \eqref{linkage2} with $\beta=1$. The thick solid lines numerical \eqref{U exact} for both Janus \eqref{Janus gen} activity distribution and the generalized Janus \eqref{Janus gen}. Also shown are the corresponding small-Damk\"ohler-number \eqref{regular}. %With ${\mathcal{S}}$ comparable to $\mathcal B$ the small-${\mathcal{S}}$ approximation holds well at small $\mathcal B$.

In calculating $U$ using  \eqref{U exact}, we have encountered difficulties when applying the numerical scheme at large values of $\mathcal{S}$. These 
are more pronounced for the Janus profile \eqref{Janus}, where the interfacial activity undergoes a finite discontinuity at $\theta=\pi/2$. %intensified for the generalized Janus profile \eqref{Janus gen}. 
Apparently, the associated non-smoothness escalates the Gibbs phenomenon. In any event,
  the approach to the limit \eqref{large Da_b U linked Janus 2} is unequivocal.
\section{Concluding remarks}  \label{sec:conclusion}
We have analyzed  self-phoresis of active colloids in situations where solute is transported by diffusion and consumed by two chemical reactions, one at the colloid boundary and one within the bulk. The dimensionless problem is governed by the two 
associated Damk\"ohler numbers, $\mathcal S$ and $\mathcal B$. %The dependence of these numbers upon the dimensional quantities in the problem specification suggests two possible natural linkages between $\mathcal S$ and $\mathcal B$. 

We have solved the problem using an eigenfunction expansion. This semi-analytic solution, applicable for all values of $\mathcal S$ and $\mathcal B$, has been accompanied by asymptotic approximations. The chosen limits correspond to two  possible natural linkages between $\mathcal S$ and $\mathcal B$. 
For small Damk\"ohler numbers, the particle velocity is proportional to $\mathcal S$ and independent of $\mathcal B$. %It effectively coincides with the velocity calculated in the flux-prescribed model of Golestanian \textit{et al.} \cite{Golestanian:07}.
At large Damk\"ohler numbers, the solute concentration is uniform except within a boundary layer about the active portion of the boundary. 
The details of the boundary-layer transport depend upon the linkage between $\mathcal S$ and $\mathcal B$. In particular, for $\mathcal S\propto\mathcal B$ we find that the particle velocity depends upon the relative fraction of the active boundary, but is otherwise  indifferent to the activity details in that fraction. The associated boundary-layer solution breaks down near the edge of the active portion of the boundary. Following a similar analysis in a classical wave problem \cite{Sommerfeld:96}, we have obtained a close-form solution of the local transport problem in the edge region.

%(A similar discontinuity for linkage by size when $f$ is discontinuous at that boundary, see indeed \eqref{A case I}.)

We summarize our results in terms of dimensional quantities. For weak chemical activity, the particle velocity is proportional to $b\bar{k}c_\infty/D$. %, with proportionality coefficient function of activity profile. 
This size-independent scaling %, which is independent of linkage, 
is the same as in the simplest models of flux-prescribed distributions \citep{Golestanian:07}, the flux scale being $\bar{k}c_\infty$.
In the limit of strong activity, the velocity scales as $bc_\infty/a$. Here, there are two situations. If the limit is realized by large values of $a$, the ratio of the particle velocity to $bc_\infty/a$ depends (nonlinearly) upon both the ratio $\alpha=\bar{k}/\sqrt{k_bD}$ and the activity profile. If the limit is realized by small values of $D$, the ratio of the particle velocity to $bc_\infty/a$ is independent of the reaction coefficients, the solute diffusivity, and even the details of the reaction profile. 
\appendix*
\section{Transition region}
Following \cite{Lamb:07}, we seek a solution of \eqref{modified Helmholtz} of the form
\begin{equation}
C = \ee^{-Y} G + \ee^Y H. \label{trick}
\end{equation}
Requiring the functions $G$ and $H$ to satisfy
\begin{equation}
\pd{^2G}{X^2}+\pd{^2G}{Y^2} = 2\pd{G}{Y}, \quad
\pd{^2H}{X^2}+\pd{^2H}{Y^2} = -2\pd{H}{Y},\label{G H}
\end{equation}
\eqref{modified Helmholtz} is trivially satisfied.
To solve equations \eqref{G H}, we employ the parabolic-cylinder coordinates (see Fig.~\ref{fig:schematic})
\begin{equation}
\xi = \rho^{1/2} \cos\frac{\vartheta}{2}, \quad \eta =  \rho^{1/2} \sin\frac{\vartheta}{2}.
\label{parabolic-cylinder}
\end{equation}
These are natural for the transition-region geometry and conditions \eqref{transition left BC}--\eqref{simple BC}, since the negative real axis becomes
$\xi=0$, while the positive real axis becomes $\eta=0$. 
With $\xi$ and $\eta$ as independent variables,  \eqref{G H} become
\refstepcounter{equation}
$$
\pd{^2G}{\xi^2}+\pd{^2G}{\eta^2} = 4\left(\eta\pd{G}{\xi} + \xi\pd{G}{\eta}\right), \quad
\pd{^2H}{\xi^2}+\pd{^2H}{\eta^2} = -4\left(\eta\pd{H}{\xi} + \xi\pd{H}{\eta}\right).
\label{G H better}
\eqno{(\theequation{\mathit{a},\mathit{b}})}
$$

The solution to (\ref{G H better}a) can be written as a combination of two similarity solutions,
\begin{equation}
G = G_+(\zeta_+) + G_-(\zeta_-),
\end{equation}
wherein $\zeta_\pm = \xi\pm\eta$. We therefore obtain the ordinary differential equations,
\begin{equation}
G_+'' = 2\zeta_+ G_+', \quad G_-'' = -2\zeta_- G_-',
\end{equation}
which integrate to give $G_\pm' = g_\pm \ee^{\pm \zeta_\pm^2}$.
Similarly, the solution to (\ref{G H better}b) is written as a combination of two similarity solutions,
\begin{equation}
H = H_+(\zeta_+) + H_-(\zeta_-).
\end{equation}
The resulting equations,
\begin{equation}
H_+'' = -2\zeta_+ H_+', \quad H_-'' = 2\zeta_- H_-',
\end{equation}
integrate to give $H_\pm' = h_\pm \ee^{\mp \zeta_\pm^2}$.

Now, as $\rho\to\infty$, it is evident that $\zeta_+\to\infty$ for all $0<\vartheta<\pi$ while $\zeta_-$ tends to $\infty$ for  $0<\vartheta<\pi/2$ and to $-\infty$ for  $\pi/2<\vartheta<\pi$. To avoid a super-exponential divergence of $C$ at large $\rho$, which would clearly contradict \eqref{transition decay}, we must set
$g_+ = h_-=0$. We conclude that the most general solutions of \eqref{G H better} are
\begin{equation}
G(\xi,\eta) = \grave g + g \erf(\xi-\eta), \quad H(\xi,\eta) = \grave h + h \erf(\xi+\eta). \label{only A B M N}
\end{equation}
The four constants appearing in \eqref{only A B M N} are determined from the boundary conditions. With condition \eqref{transition left BC} applying at $\xi=0$ we readily obtain $\grave h=\grave g$ and $h=-g$. Thus,
\eqref{trick} and \eqref{only A B M N} give
\begin{equation}
C = \ee^{-Y} [\grave g+g\erf(\xi-\eta)] + \ee^Y [\grave g-g\erf(\xi+\eta)].
\end{equation}
Recalling that $\erf z \sim 1 - \ee^{-z^2}/z\sqrt{\pi}$ for $z\to\infty$, we must impose
$\grave g=g$ to %prevent an exponential divergences of $C$    at large $Y$, which would otherwise contradict 
satisfy condition \eqref{transition decay}. Last, noting that the inhomogeneous condition \eqref{simple BC} applies at $\eta=0$, we readily obtain $g=1/2$. We conclude that 
\begin{equation}
C = \frac{\ee^{-Y}}{2} \left[ 1 + \erf (\xi-\eta)\right] + 
\frac{\ee^{Y}}{2} \left[ 1 - \erf (\xi+\eta) \right].
\end{equation}
Substitution of \eqref{parabolic-cylinder} yields \eqref{transition final}.
\bibliography{refs}
\end{document}